\documentstyle[amssymb,12pt,thmsa,sw20lart,epsfig]{article}

\input{tcilatex}
\begin{document}

\pagestyle{empty}

\begin{center}

\begin{tabular}{c}
\end{tabular}

\vskip 2cm

ON WARD-TAKAHASHI IDENTITIES FOR THE PARISI SPIN GLASS

\vskip 3cm

C. De Dominicis$^{\text{a}}$, T. Temesvari$^{\text{b}}$, I.\ Kondor$^{%
\text{c,d}}$

\vskip 1cm

$^{\text{a}}$Service de Physique Th\'{e}orique

CEA/Saclay

F-91191 Gif-sur-Yvette Cedex

FRANCE

(cirano@spht.saclay.cea.fr)

\vskip 0.5cm

$^{\text{b}}$Department of Theoretical Physics

E\"{o}tv\"{o}s University, Puskin u. 5/7

H-1088 Budapest

HONGRIE

(temtam@hal9000.elte.hu)

\vskip 0.5cm

$^{\text{c}}$Department of Complex Systems

E\"{o}tv\"{o}s University, Muzeum krt 6/8

H-1088 Budapest

HONGRIE

\vskip 0.5cm

$^{\text{d}}$Bolyai Collegium

E\"{o}tv\"{o}s University, Amerikai ut 96

H-1088 Budapest

HONGRIE

(kondori@ludens.elte.hu)
\end{center}
\eject
{
\parindent=1.5cm\narrower
\noindent{\bf Abstract.} The introduction of ``small permutations'' allows us to
derive Ward-Takahashi identities for the spin-glass, in the Parisi limit of
an infinite number of steps of replica symmetry breaking.\ The first
identities express the emergence of a band of Goldstone modes 
. The next
identities relate components of (the Replica Fourier Transformed) 3-point
function to overlap derivatives of the 2-point function (inverse
propagator).\ A jump in this last function is exhibited, when its two
overlaps are crossing each other, in the special simpler case where one of
the cross-overlaps is maximal.\par}

\vskip 1cm

\centerline {This work is a tribute to the memory of Giovanni Paladin}

\section{Introduction and summary}

The breaking of a continuous symmetry group is known to generate
massless or Goldstone [1] modes in the broken symmetry phase.\ Infrared
divergences associated with massless-modes usually complicate the
renormalization process inside that phase. However the broken invariance
generates Ward-Takahashi (W-T) identities, the first of which imposes
masslessness. Higher order identities are then instrumental in the taming of
proliferating divergences into relationships between renormalized quantities
[2, 3].

For systems with quenched disorder whose broken invariance group is the
permutation group of replicas, the affair is more subtle. The replica
symmetry may undergo $R$ steps of breaking [4]. For $R=1$ (or $R$ finite)
there will be no Goldstone mode.\ For the Parisi limit [4] $R\rightarrow
\infty ,$ one finds bands of massless modes [5, 6, 7] and highly singular
bare propagators [8]. Thus the need for help from W-T identities is even
more acute in this case.

In a separate publication [9] hereafter called (I) the basic approach to a
derivation of W-T identities was presented with the introduction of ``small
permutations'' (the small parameter being e.g. $p_{r}-p_{r+1}\simeq 1/R,$
the size difference between consecutive Parisi boxes). There, the result for
the first W-T identities, that, in particular, exhibit Goldstone modes, was
presented. In this work we shall follow the same approach.\ In the Parisi
limit, invariance properties, associated with the replica permutation group,
become hidden, and a new invariance, the so called reparametrization
invariance, emerges. Their interrelationship, examined in (I) will not be
further investigated here.

In Section 2 we recall briefly relationships leading to W-T identities in
the continuum limit. Parametrization of replicas and ``small permutations''
are defined in Section 3. We recall the 2-point function parametrization in
Section 4. The first W-T identity, yielding a continuity condition, is
derived in Section 5. The W-T identity, imposing masslessness is derived in
Section 6. Section 7 is devoted to a minimal discussion of the
parametrization for 3-point functions.\ Finally sections 8--10 contain the
derivation of W-T identities relating 3-point and 2-point functions.\ To
keep the developments to a reasonable size, while still giving a detailed
derivation, we have chosen the simpler case where one of the cross-overlaps
is maximal (and equal to $R+1).$

\section{Invariance properties and W-T relationships:}

Let us consider the free energy functional $F\left\{ \bar{q}\right\} $
where $\bar{q}$ is the $n\left( n-1\right) /2$ dimensional order parameter
vector $q_{\alpha \beta }$ (with $\alpha =1,2...n;\,q_{\alpha \beta
}=q_{\beta \alpha };\,q_{\alpha \alpha }=0).$ Let $\bar{h}$ be a general
source conjugate to $\bar{q},F$ being the Legendre transform of $W\left\{ 
\bar{h}\right\} \equiv \ln Z$

\begin{equation}
W\left\{ \bar{h}\right\} +F\left\{ \bar{q}\right\} =\bar{h}\cdot \bar{q}=%
\tfrac{1}{2}\sum_{\alpha ,\beta }h_{\alpha \beta }q_{\alpha \beta } 
\tag{2.1}
\end{equation}

By construction $W$ and $F$ are invariant under a replica permutation $P,$
which is also a permutation i.e. a rotation of the $n\left( n-1\right) /2$
axes, changing $\bar{q}$ into $\bar{q}^{\prime }$

\begin{equation}
\bar{q}^{\prime }=P\bar{q}  \tag{2.2}
\end{equation}

\begin{equation}
Pq_{\alpha \beta }=q_{P_{\alpha };P_{\beta }}  \tag{2.3}
\end{equation}
We thus have from replica permutation invariance,

\begin{equation}
F\left\{ \bar{q}^{\prime }\right\} =F\left\{ \bar{q}\right\}  \tag{2.4}
\end{equation}
From (2.1) we also have

\begin{equation}
\frac{\delta F\left\{ \bar{q}\right\} }{\delta q_{\alpha \beta }}=h_{\alpha
\beta }  \tag{2.5}
\end{equation}
and the identity

\begin{equation}
\frac{\delta F\left\{ \bar{q}^{\prime }\right\} }{\delta q_{\alpha \beta }}%
=h_{\alpha \beta }^{\prime }.  \tag{2.6}
\end{equation}
This last equation states that in (2.5) it is equivalent to rotate the
source $\bar{h}$ or to rotate the order parameter $\bar{q},$ $\partial
F/\partial \bar{q}$ is functionally depending upon. Separating out in $P,$
the identity operator

\begin{equation}
P\equiv 1+\delta P  \tag{2.7}
\end{equation}

\begin{equation}
\delta Pq_{\alpha \beta }=q_{P_{\alpha },P_{\beta }}-q_{\alpha \beta } 
\tag{2.8}
\end{equation}
we have

\begin{equation}
\frac{\partial F}{\partial \bar{q}}\left\{ \bar{q}+\delta P\bar{q}\right\} =%
\bar{h}+\delta P\bar{h}  \tag{2.9}
\end{equation}
i.e. by Taylor expanding around $\bar{q}$

\begin{equation}
\frac{1}{2}\sum_{\mu \nu }\frac{\partial ^{2}F}{\partial q_{\alpha \beta
}\partial q_{\mu \nu }}\left\{ \bar{q}\right\} \delta Pq_{\mu \nu
}+...=\delta Ph_{\alpha \beta }  \tag{2.10}
\end{equation}
Under a ``small permutation'', one can neglect higher order terms in (2.10),
and recover the first W-T identity relating 2-point and 1-point functions.
Applying the same procedure (2.6--9) to $\partial ^{2}F/\partial q_{\alpha
\beta }\partial q_{\gamma \delta }$ will produce the next set of identities
relating 3-point and 2-point functions.

\section{``Small permutations$^{1}$'':}

\footnotetext[1]{
The first attempt at building an ``infinitesimal permutation'' can be found
in ref.[10].\ A general discussion of these transformations has been given
by Goltsev[11].\ Independently Parisi and Slanina[12] have introduced
similar constructions in a random polymer context.}

Let us parametrize a replica $\alpha $ by its address, i.e. the list
of branch numbers

\begin{equation}
\alpha :\left[ a_{{\rm o}},a_{1},a_{2}...a_{R}\right]  \tag{3.1}
\end{equation}
one has to follow to reach replica $\alpha $ at the bottom of the
(ultrametric) tree.\ Here

\begin{eqnarray}
a_{{\rm o}} &=&0,1,2,...\left( p_{{\rm o}}/p_{1}-1\right)  \nonumber \\
&&  \nonumber \\
a_{1} &=&0,1,2,...\left( p_{1}/p_{2}-1\right)  \tag{3.2}
\end{eqnarray}
and $a_{r}$ is numbering the $p_{r}/p_{r+1},$ branches\footnote[2]{%
We could have also numbered $a_{{\rm o}}=1,2,...p_{{\rm o}}/p_{1}.$ One
should think of $a_{{\rm o}}$ as taking values on a circle i.e. mod $\left(
p_{{\rm o}}/p_{1}\right) .$} descending from a node of level $r.$ Two
replicas $\alpha ,\beta $ have an overlap $r$

\begin{equation}
\alpha \cap \beta =r  \tag{3.3}
\end{equation}
if

\begin{equation}
\begin{array}{l}
\alpha \\ 
\\ 
\beta
\end{array}
\left[ 
\begin{array}{ll}
& a_{r}a_{r+1}...a_{R} \\ 
a_{{\rm o}}a_{1}...a_{r-1} &  \\ 
& b_{r}b_{r+1}...b_{R}
\end{array}
\right] \qquad \qquad \qquad a_{r}\neq b_{r}  \tag{3.4}
\end{equation}

When $q_{\alpha \beta }$ takes its saddle-point value, we then have

\begin{equation}
q_{\alpha \beta }=q_{r}  \tag{3.5}
\end{equation}

We now define a ``small permutation'' $P^{\left( r\right) }$ by its action
upon the addresses.\ For example, for all replicas with a given fixed $%
a_{r+1},$ say

\begin{equation}
a_{r+1}=0_{r+1}  \tag{3.6}
\end{equation}
{\it and only for those}, the action of $P^{\left( r\right) }$ will be to
change $a_{r}$ into $1+a_{r}.$ For all other replicas, $P^{\left( r\right) }$
will act as the identity operator,

\begin{eqnarray}
P_{\alpha }^{\left( r\right) } &\equiv &P^{\left( r\right) }\left[ a_{{\rm o}%
},a_{1}...a_{r},a_{r+1},...a_{R}\right]  \nonumber \\
&=&\left[ a_{{\rm o}},a_{1}...a_{r},a_{r+1},...a_{R}\right] \left( 1-\delta
_{a_{r+1};{\rm o}_{r+1}}\right) +\left[ a_{{\rm o}%
},a_{1},...1+a_{r},a_{r+1},...a_{R}\right]  \nonumber \\
&&\delta _{a_{r+1};{\rm o}_{r+1}}  \tag{3.7}
\end{eqnarray}
This choice of permutation will be kept throughout the paper. In words, what
our chosen $P^{\left( r\right) }$ does is the following: consider, in the
ultrametric tree, a given node at level $r\,$and its $p_{r}/p_{r+1}$ nodes
at level $r+1$ descending from it. From each one of these $p_{r}/p_{r+1}$
nodes emerge $p_{r+1}/p_{r+2}$ branches. Select the 0th one (and its
descendent) and the set of replicas it leads to.\ We have $p_{r}/p_{r+1}$
such sets.\ $P^{\left( r\right) }\,$circularly permutes {\it en bloc}, those 
$p_{r}/p_{r+1}$ sets of replicas.\ What we have just said for one given node
at level $r,$ is valid for all such $p_{{\rm o}}/p_{r}$ nodes.

Let us compute now the action of $P^{\left( r\right) }$ upon $q_{\alpha
\beta },$ i.e.

\begin{equation}
\delta P^{\left( r\right) }q_{\alpha \beta }=q_{P_{\alpha }^{\left( r\right)
};P_{\beta }^{\left( r\right) }}-q_{\alpha \beta .}  \tag{3.8}
\end{equation}
If $\alpha \cap \beta =r+1,$ i.e. for $b_{r+1}\neq 0_{r+1},$ we have

\begin{eqnarray}
\delta P^{\left( r\right) }q_{\alpha \beta } &=&q\left[ 
\begin{array}{lll}
& 1+a_{r} & 0_{r+1}...a_{R} \\ 
a_{{\rm o}}...a_{r-1} &  &  \\ 
& \quad \,a_{r} & b_{r+1}...b_{R}
\end{array}
\right] -q\left[ 
\begin{array}{lll}
& a_{r} & 0_{r+1}...a_{R} \\ 
a_{{\rm o}}...a_{r-1} &  &  \\ 
& a_{r} & b_{r+1}...b_{R}
\end{array}
\right]  \nonumber \\
&=&q_{r}-q_{r+1}  \tag{3.9}
\end{eqnarray}
If $\alpha \cap \beta =r,$ we have, if and only if $b_{r+1}\neq 0_{r+1}$

\begin{eqnarray}
\delta P^{\left( r\right) }q_{\alpha \beta } &=&q\left[ 
\begin{array}{lll}
& a_{r} & 0_{r+1}...a_{R} \\ 
a_{{\rm o}}...a_{r-1} &  &  \\ 
& a_{r} & b_{r+1}...b_{R}
\end{array}
\right] -q\left[ 
\begin{array}{lll}
& -1+a_{r} & 0_{r+1}...a_{R} \\ 
a_{{\rm o}}...a_{r-1} &  &  \\ 
& \quad \;\,a_{r} & b_{r+1}...b_{R}
\end{array}
\right]  \nonumber \\
&=&q_{r+1}-q_{r}.  \tag{3.10}
\end{eqnarray}

and likewise under $\alpha ,\beta $ exchange.\ For {\it all other components 
}$q_{\alpha \beta },$ we have

\begin{equation}
\delta P^{\left( r\right) }q_{\alpha \beta }=0  \tag{3.11}
\end{equation}
When $R$ is large (and in particular in the Parisi limit) we obtain (except
for $r=R),$

\begin{equation}
q_{r}-q_{r+1}\sim p_{r}-p_{r+1}\sim O\left( 1/R\right)  \tag{3.12}
\end{equation}
allowing us to discard higher order terms in the Taylor expansion (2.9). We
can then write

\begin{equation}
\frac{1}{2}\sum_{\mu \nu }\frac{\partial ^{2}F}{\partial q_{\alpha \beta
}\partial q_{\mu \nu }}\delta P^{\left( r\right) }q_{\mu \nu }=\delta
P^{\left( r\right) }h_{\alpha \beta }+O\left( 1/R\right)  \tag{3.13}
\end{equation}
\begin{equation}
\frac{1}{2}\sum_{\mu \nu }\frac{\partial ^{3}F}{\partial q_{\alpha \beta
}\partial q_{\gamma \delta }\partial q_{\mu \nu }}\delta P^{\left( r\right)
}q_{\mu \nu }=\delta P^{\left( r\right) }\frac{\delta ^{2}F}{\partial
q_{\alpha \beta }\partial q_{\gamma \delta }}+O\left( 1/R\right)  \tag{3.14}
\end{equation}
where, in the right side of this equation $\delta P^{\left( r\right) }$ acts
upon $\alpha \beta \gamma \delta .$

We have now to implement these relationships using explicit parametrizations
for the two and three-point functions, namely

\begin{equation}
\frac{\partial ^{2}F}{\partial q_{\alpha \beta }\partial q_{\mu \nu }}\equiv
M^{\alpha \beta ;\mu \nu }.  \tag{3.15}
\end{equation}
the ``mass-operator'' (inverse propagator), and

\begin{equation}
\frac{\partial ^{3}F}{\partial q_{\alpha \beta }\partial q_{\gamma \delta
}\partial q_{\mu \nu }}\equiv W^{\alpha \beta ;\gamma \delta ;\mu \nu }. 
\tag{3.16}
\end{equation}

\section{2-point function parametrization:}

We give now a minimal analysis of the 2-point function $M^{\alpha
\beta ;\gamma \delta },$ i.e. the (zero momentum) inverse propagator.
Details can be found in [8] and for the Replica Fourier Transform (RFT)
approach in [13].

With four replicas, $M^{\alpha \beta ;\gamma \delta }$ will depend upon
three overlaps.

{\it (i)} If generically $\alpha \cap \beta \equiv \gamma \cap \delta =r,$
(Replicon configurations), $M$ depends upon two cross-overlaps $u,v\geqslant
r+1:$

\begin{equation}
M_{u;v}^{r;r}\qquad 
\begin{array}{l}
u=\max \left( \alpha \cap \gamma ,\alpha \cap \delta \right) \\ 
v=\max \left( \beta \cap \gamma ,\beta \cap \delta \right)
\end{array}
\tag{4.1}
\end{equation}
The double RFT defines the Replicon kernel

\begin{equation}
_{R}M_{\widehat{k},\widehat{\ell }}^{r;r}=\sum_{u=k}^{R+1}\sum_{v=\ell
}^{R+1}p_{u}p_{v}\left(
M_{u;v}^{r;r}-M_{u-1;v}^{r;r}-M_{u;v-1}^{r;r}+M_{u-1,v-1}^{r;r}\right) 
\tag{4.2}
\end{equation}
for $k,\ell \geqslant r+1.$ Hatted variables stand for RFT ones, this
notation is preferred here to the one that uses a different symbol for
kernels (i.e. RFT's), so that we have the possibility of mixed (or
incomplete) transforms.

The Replicon component $_{R}M_{u;v}^{r;r}$ is in turn the inverse double
transform

\begin{equation}
_{R}M_{u;v}^{r;r}=\sum_{k=r+1}^{u}\sum_{\ell =r+1}^{v}\frac{1}{p_{k}}\frac{1%
}{p_{\ell }}\left( M_{\widehat{k},\widehat{\ell }}^{r;r}-M_{\widehat{k+1};%
\hat{\ell}}^{r;r}-M_{\widehat{k};\widehat{\ell +1}}^{r;r}+M_{\widehat{k+1};%
\widehat{\ell +1}}^{r;r}\right) .  \tag{4.3}
\end{equation}

{\it (ii)} If generically $\alpha \cap \beta =r,\gamma \cap \delta =s,r\neq
s $ (Longitudinal-Anomalous configurations), $M$ depends upon one
cross-overlap

\begin{equation}
_{A}M_{t}^{r;s}\equiv M_{t}^{r;s}\qquad \qquad \qquad \qquad t=\max \left(
\alpha \cap \gamma ,\alpha \cap \delta ,\beta \cap \gamma ,\beta \cap \delta
\right)  \tag{4.4}
\end{equation}
and

\begin{equation}
M_{\widehat{k}}^{r;s}=\sum_{t=k}^{R+1}p_{t}^{\left( r;s\right) }\left(
M_{t}^{r;s}-M_{t-1}^{r;s}\right)  \tag{4.5}
\end{equation}

\begin{equation}
M_{t}^{r;s}=\sum_{k={\rm o}}^{k=t}\frac{1}{p_{k}^{\left( r;s\right) }}\left(
M_{\widehat{k}}^{r;s}-M_{\widehat{k+1}}^{r;s}\right)  \tag{4.6}
\end{equation}
with, e.g. for $r<s,$

\begin{equation}
p_{x}^{\left( r;s\right) }= 
\begin{array}{ll}
p_{x} & x\leqslant r<s \\ 
2p_{x} & r<x\leqslant s \\ 
4p_{x} & r<s<x
\end{array}
\tag{4.7}
\end{equation}

The description is then complete if one gives $_{A}M$ in the Replicon
configuration with $r\equiv s,$ and $u,v\geqslant r+1,$

\begin{equation}
_{A}M_{u;v}^{r;r}=_{A\!}M_{u}^{r;r}+_{A\!}M_{v}^{r;r}-_{A\!}M_{r}^{r;r} 
\tag{4.8}
\end{equation}
a component thus projected out in the {\it double} RFT of (4.2). In all
cases we may now think of $_{A}M$ with a single lower index (hatted or not)
and $_{R}M$ with two lower indices (hatted or not).

\section{Relating 2-point and 1-point functions: $\delta P_{\alpha \beta
}^{\left( r\right) }=0$}

We explicit now eq.(3.13), where $\alpha $ and $\beta $ can be chosen
at our convenience. The simplest case occurs for $\delta P_{\alpha \beta
}^{\left( r\right) }=0,$ hence with a {\it zero right hand side.} For example%
\footnote[3]{%
with $P^{\left( r\right) }$ as of (3.7) the choice $\alpha \cap \beta =r+1$
leads to a slightly simpler calculation than $\alpha \cap \beta =r.$} if $%
\alpha \cap \beta =r+1,$ i.e. $a_{r+1}\neq b_{r+1},$ we have

\begin{equation}
\begin{array}{l}
\alpha \\ 
\\ 
\beta
\end{array}
\left[ 
\begin{array}{lll}
& a_{r} & a_{r+1}...a_{R} \\ 
a_{{\rm o}}...a_{r-1} &  &  \\ 
& a_{r} & b_{r+1}...b_{R}
\end{array}
\right]  \tag{5.1}
\end{equation}
and, if we then choose

\begin{equation}
a_{r+1},b_{r+1}\neq {\rm 0}_{r+1}  \tag{5.2}
\end{equation}
we have $\delta P^{\left( r\right) }h_{\alpha \beta }=0,$ i.e. a null right
hand side.

We collect now all the non vanishing contributions coming up in the sum over 
$\mu ,\nu $ in (3.13). The factor $\delta P^{\left( r\right) }q_{\mu \nu }$
is non zero and equal to $\pm \left( q_{r}-q_{r+1}\right) $ when $\delta
P^{\left( r\right) }$ acts upon

\begin{equation}
\qquad \; 
\begin{array}{l}
\mu \\ 
\\ 
\nu
\end{array}
\left[ 
\begin{array}{lll}
& m_{r} & 0_{r+1}...m_{R} \\ 
m_{{\rm o}}...m_{r-1} &  &  \\ 
& m_{r} & n_{r+1}^{\prime }...n_{R}
\end{array}
\right] \qquad \qquad \qquad \;\;\mu \cap \nu =r+1  \tag{5.3}
\end{equation}
or

\begin{equation}
\qquad \; 
\begin{array}{l}
\mu \\ 
\\ 
\nu
\end{array}
\left[ 
\begin{array}{lll}
& -1+m_{r} & 0_{r+1}...m_{R} \\ 
m_{{\rm o}}...m_{r-1} &  &  \\ 
& \quad \;m_{r} & n_{r+1}^{\prime }...n_{R}
\end{array}
\right] \qquad \qquad \mu \cap \nu =r\;\;\;  \tag{5.4}
\end{equation}

respectively (see eq.3.9-10).\ Here a primed component is distinct from zero.

Let us note as a preliminary simplifying remark, that the sum over $\mu ,\nu 
$ gives a weight $\left( p_{r+1}/p_{r+2}-1\right) \simeq O\left( 1/R\right) $
to all contributions where the sum can be freely carried out, ignoring the
passive replicas (here $\alpha ,\beta ,\alpha \cap \beta \!=\!r\!+\!1).$ As
a consequence only $\mu \nu $ configurations with $\left( m_{{\rm o}}=a_{%
{\rm o}},m_{1}=a_{1}...,m_{r}=a_{r}\right) $ need be considered.

The $\mu ,\nu $ sum, i.e. the sum over $m^{\prime }s$ and $n^{\prime }s,$ is
thus reduced to two types of geometries

{\it (i)} $\mu \cap \nu =r+1$

\begin{equation}
\begin{array}{l}
\mu \\ 
\\ 
\nu
\end{array}
\left[ 
\begin{array}{lll}
& a_{r} & 0_{r+1}...m_{R} \\ 
a_{{\rm o}}...a_{r-1} &  &  \\ 
& a_{r} & n_{r+1}^{\prime }...n_{R}
\end{array}
\right]  \tag{5.5}
\end{equation}

{\it (ii) }$\mu \cap \nu =r$

\begin{equation}
\begin{array}{l}
\mu \\ 
\\ 
\nu
\end{array}
\left[ 
\begin{array}{lll}
& -1+a_{r} & 0_{r+1}...m_{R} \\ 
a_{{\rm o}}...a_{r-1} &  &  \\ 
& \quad \;\,a_{r} & n_{r+1}^{\prime }...n_{R}
\end{array}
\right]  \tag{5.6}
\end{equation}
plus $\mu ,\nu $ exchange.

In case (i) we obtain a contribution

\begin{equation}
\left[ M_{r+1}^{r+1;r+1}\left( \frac{p_{r+1}}{p_{r+2}}-3\right)
p_{r+2}^{2}+2\sum_{u=r+2}^{R+1}M_{u}^{r+1;r+1}\left( p_{u}-p_{u+1}\right)
p_{r+2}\right] \left( q_{r}-q_{r+1}\right)  \tag{5.7}
\end{equation}
Here the first term is for $n_{r+1}^{\prime }\neq a_{r+1},b_{r+1},0_{r+1}$
i.e. taking $\left( p_{r+1}/p_{r+2}-3\right) $ values, $p_{r+2}^{2}$ coming
from summation over free branch numbers $m_{r+2},m_{r+3}...$ and $%
n_{r+2},n_{r+3}...$The second term arises from $n_{r+1}^{\prime
}=a_{r+1}\left( {\rm or\ }b_{r+1}\right) $ and summing over free branch
numbers when $\nu \cap \alpha =u,\;u\geqslant r+2\;\left( {\rm or\ }\nu \cap
\beta =u\right) .$ Indeed the sum is then over $n_{u}\neq a_{u},$ and $%
n_{u+1},n_{u+2},...$ that is yielding $\left( p_{u}/p_{u+1}-1\right)
p_{u+1}. $

In case (ii) the contribution becomes instead

\begin{equation}
\left[ M_{r+1}^{r+1;r}\left( \frac{p_{r+1}}{p_{r+2}}-3\right)
p_{r+1}^{2}+2\sum_{u=r+2}^{R+1}M_{u}^{r+1;r}\left( p_{u}-p_{u+1}\right)
p_{r+2}\right] \left( q_{r+1}-q_{r}\right)  \tag{5.8}
\end{equation}
Using now the obvious relationship,

\begin{equation}
\sum_{u=r+2}^{R+1}M_{u}^{r+1;s}\left( p_{u}-p_{u+1}\right)
=\sum_{u=r+2}^{R+1}p_{u}\left( M_{u}^{r+1;s}-M_{u-1}^{r+1;s}\right)
+p_{r+2}M_{r+1}^{r+1;s}  \tag{5.9}
\end{equation}
and the RFT definitions (4.5, 7), then, the sum of (5.7) and (5.8) becomes

\begin{equation}
\left[ \frac{p_{r+2}}{4}\left( M_{\widehat{r+2}}^{r+1;r+1}-M_{\widehat{r+2}%
}^{r+1;r}\right) +O\left( 1/R\right) \right] \left( q_{r}-q_{r+1}\right) =0 
\tag{5.10}
\end{equation}
This first relationship expresses, in the $R\rightarrow \infty $ limit, the
continuity of the kernels (Fourier Transforms) in their overlaps.

\section{Relating 2-point and 1-point functions: $\delta P_{\alpha \beta
}^{\left( r\right) }\neq 0$}

Let us choose again $\alpha \cap \beta =r+1$ but now with $\delta
P^{\left( r\right) }q_{\alpha \beta }=q_{r}-q_{r+1}.$ Hence the right hand
side of (3.13) is now

\begin{equation}
\left( h_{r}-h_{r+1}\right) .  \tag{6.1}
\end{equation}
and we have

\begin{equation}
\begin{array}{l}
\alpha \\ 
\\ 
\beta
\end{array}
\left[ 
\begin{array}{lll}
& a_{r} & 0_{r+1}...a_{R} \\ 
a_{{\rm o}}...a_{r-1} &  &  \\ 
& a_{r} & b_{r+1}^{\prime }...b_{R}
\end{array}
\right] .  \tag{6.2}
\end{equation}

Carrying out again the $\mu ,\nu $ sum in (3.13) we have the two cases
considered above, all the others yielding zero or a contribution of order $%
1/R$ as remarked before.

(i) $\mu \cap \nu =r+1$ as in (5.5). We get

\begin{equation}
\left[ \sum_{u=r+2}^{R+1}M_{u}^{r+1;r+1}\delta u\left( \frac{p_{r+1}}{p_{r+2}%
}-2\right)
p_{r+2}+\sum_{u=r+2}^{R+1}\;\sum_{v=r+2}^{R+1}M_{u;v}^{r+1;r+1}\delta
u\;\delta v\right] \left( q_{r}-q_{r+1}\right)  \tag{6.3}
\end{equation}
Here we use $\delta u\equiv p_{u}-p_{u+1}.$

The first term in(6.3) comes from $n_{r+1}^{\prime }\neq b_{r+1}^{\prime
},0_{r+1}$ i.e. taking $\left( p_{r+1}/p_{r+2}-2\right) $ values, the factor 
$p_{r+2}$ from summing over $m_{r+2},...,m_{R},$ and $\delta u$ from summing
over $n_{u}\neq a_{u},n_{u+1}...,n_{R}$ when $\gamma \cap \alpha
=u,\;u\geqslant r+2.$ The last term is from $n_{r+1}^{\prime
}=b_{r+1}^{\prime }$ and summing over free branch numbers $m$ and $n,\;\mu
\cap \alpha =u,\;\gamma \cap \beta =v,\;u,v\geqslant r+2$ (and $\mu ,\nu $
exchange).

(ii)$\mu \cap \nu =r$ as in (5.6), we get instead

\begin{equation}
\left[ M_{r+1}^{r+1;r}\left( \frac{p_{r+1}}{p_{r+2}}-2\right)
p_{r+2}^{2}+\sum_{v=r+2}^{R+1}M_{v}^{r+1;r}\delta v\;p_{r+2}\right] \left(
q_{r+1}-q_{r}\right)   \tag{6.4}
\end{equation}
where the first term is for $n_{r+1}^{\prime }\neq b_{r+1}^{\prime },0_{r+1}$
and the last for $\nu \cap \beta =v,\,v\geqslant r+2$ and the resulting sum
over free branch numbers $m$ and $n.$

Pulling (6.1, 3.5) together, and using relationships of eq.(5.9), and

\begin{eqnarray}
\sum_{u=r+2}^{R+1}\;\sum_{v=r+2}^{R+1}M_{u;v}^{r+1;r+1}\;\delta u\;\delta v
&=&M_{\widehat{r+2};\widehat{r+2}}^{r+1;r+1}+2p_{r+2}%
\sum_{u=r+2}^{R+1}M_{u}^{r+1;r+1}\delta u-p_{r+2}^{2}M_{r+1}^{r+1;r+1} 
\nonumber \\
&&  \tag{6.5}
\end{eqnarray}
that follows from RFT definitions (4.2) and from (4.8), we obtain,

\begin{eqnarray}
&&M_{\widehat{r+2};\widehat{r+2}}^{r+1;r+1}+p_{r+2}\stackunder{u=r+2}{%
\stackrel{R+1}{\sum }}p_{u}\left[ \left(
M_{u}^{r+1;r+1}-M_{u-1}^{r+1;r+1}\right) -\left(
M_{u}^{r+1;r}-M_{u-1}^{r+1;r}\right) \right]  \nonumber \\
&=&\left( h_{r+1}-h_{r}\right) /\left( q_{r+1}-q_{r}\right) +O\left(
1/R\right) .  \tag{6.6}
\end{eqnarray}
Using definitions (4.5, 7) for RFT's, we get

\begin{equation}
M_{\widehat{r+2};\widehat{r+2}}^{r+1;r+1}+\frac{p_{r+2}}{4}\left( M_{%
\widehat{r+2}}^{r+1;r+1}-M_{\widehat{r+2}}^{r+1;r}\right) =\left(
h_{r+1}-h_{r}\right) /\left( q_{r+1}-q_{r}\right) +O\left( 1/R\right) 
\tag{6.7}
\end{equation}
i.e., with the previous W-T-like relationship of equation (5.10)

\begin{equation}
M_{\widehat{r+2};\widehat{r+2}}^{r+1;r+1}=\left( h_{r+1}-h_{r}\right)
/\left( q_{r+1}-q_{r}\right) +O\left( 1/R\right) .  \tag{6.8}
\end{equation}
Taking Parisi limit, with $x=r/R+1,R\rightarrow \infty ,$ we finally, get

\begin{equation}
M_{\widehat{x+{\rm o}};\widehat{x+{\rm o}}}^{x;x}=\dot{h}\left( x\right) /%
\dot{q}\left( x\right)  \tag{6.9}
\end{equation}
This W-T identity states that one has Goldstone modes, for each overlap $x$
for which $\dot{h}\left( x\right) =0,$ that is for an overlap independent
magnetic field. In particular one has a band of Goldstone modes with or
without the presence of a magnetic field.

\section{3-point function parametrization:}

A general discussion of 3-point function parametrization would require
too much space.\ Here we give a minimal discussion and, to keep within size,
we specialize to the case where one of the given cross-overlaps is maximal,
namely $\beta \equiv \delta .$ Instead of investigating (3.14) we are now
concerned with

\begin{equation}
\frac{1}{2}\sum_{\mu \nu }W^{\alpha \beta ;\beta \gamma ;\mu \nu }\delta
P^{\left( r\right) }q_{\mu \nu }=\delta P^{\left( r\right) }M^{\alpha \beta
;\beta \gamma }+O\left( 1/R\right) .  \tag{7.1}
\end{equation}
Further we shall concentrate upon the more interesting configuration $\alpha
\cap \beta \neq \beta \cap \gamma $ in which case, the 2-point function in
the right hand side becomes

\begin{equation}
M^{\alpha \beta ;\beta \gamma }=M_{R+1}^{r;s},\qquad \qquad \qquad r\neq s 
\tag{7.2}
\end{equation}
The general 3-point function (six replicas) involves five overlaps. Because
of the choice $\beta \equiv \delta $ , one of the cross-overlaps is now $%
\beta \cap \delta \equiv \beta \cap \beta \equiv R+1.$ With the choice $%
\alpha \cap \beta \neq \beta \cap \gamma ,\;r\neq s,$ we now have only two
distinct geometries of interest

(i)${\rm \ }\mu \cap \nu \equiv \alpha \cap \beta =r,{\rm \ }$with

\begin{equation}
W^{\alpha \beta ;\beta \gamma ;\mu \nu }=W_{u;v;R+1}^{r;s;r}\qquad \qquad
\qquad \qquad u,v\geqslant r+1  \tag{7.3}
\end{equation}
where

\begin{equation}
\begin{array}{lll}
u & = & \max \left( \alpha \cap \mu ,\alpha \cap \nu \right) \\ 
v & = & \max \left( \beta \cap \mu ,\beta \cap \nu ,\gamma \cap \mu ,\gamma
\cap \nu \right)
\end{array}
\tag{7.4}
\end{equation}
the pairs $\alpha \beta ,\mu \nu $ being in a Replicon-like geometry. The
RFT writes

\begin{equation}
W_{\widehat{k},\widehat{\ell }^{\left( s\right) };R+1}^{r;s;r}\quad \qquad
\qquad \qquad \qquad \qquad k,\ell \geqslant r+1.  \tag{7.5}
\end{equation}
the $s$ superscript in $\widehat{\ell }^{\left( s\right) }$ is to remind
that for $s>r$ the double RFT is to be calculated as in (4.2) but with
weights $p_{u},p_{v}^{\left( s\right) }$ with

\begin{equation}
p_{v}^{\left( s\right) }=\QATOP{p_{v}}{2p_{v}}\quad \QATOP{v\leqslant s}{s>v}%
.  \tag{7.6}
\end{equation}
If $s<r,$ the superscript is to be forgotten, and weights $p_{u},p_{v}$ used%
\footnote[4]{%
To make a long story short, just like we have two geometries $\left(
_{R}M^{r;r},_{A}M^{r;s}\right) $ for the 2-point fuctions, we now have four
geometries $_{RR}W^{r;r;r},_{AR}W^{r;s;r},_{NR}W^{r;s;r},_{AA}W^{r;s;q},$
where $NR$ stands for Nested Replicon. In the simpler case considered here
the Replicon-like geometry hides in fact the $AR$ and $NR$ geometries, this
distinction being transferred into the extra superscript.\ See more below in
section 9.}.

(ii) $\mu \cap \nu \neq \alpha \cap \beta $ (and $\mu \cap \nu \neq \beta
\cap \gamma )$ generically, then

\begin{equation}
W^{\alpha \beta ;\beta \gamma ;\mu \nu }=_{A\!\!\!}W_{t;R+1}^{r;s;q} 
\tag{7.7}
\end{equation}

\begin{equation}
t=\max \left( \alpha \cap \mu ,\alpha \cap \nu ,\beta \cap \mu ,\beta \cap
\nu ,\gamma \cap \mu ,\gamma \cap \nu \right) .  \tag{7.8}
\end{equation}

Here the RFT

\begin{equation}
W_{\widehat{k};R+1}^{r;s;q}  \tag{7.9}
\end{equation}
is to be calculated as in (4.5) with a $p_{t}^{\left( r,s,q\right) }$
trivially generalizing (4.7).

\section{Relating 3-point and 2-point functions: $s<r$}

Let us choose again $\alpha \cap \beta =r+1,$ and $\beta \cap \gamma
=s,\,s\neq r,r+1,$ which under $P^{\left( r\right) }$ leads to $\delta
P_{\beta \gamma }^{\left( r\right) }=0{\rm .}$ Taking besides $\delta
P_{\alpha \beta }^{\left( r\right) }=0$ would yield a continuity
relationship analog to (5.10).\ Let us choose instead $\delta P_{\alpha
\beta }^{\left( r\right) }\neq 0$ as in (6.2), which gives for the right
hand side of (7.1)

\begin{equation}
\left( M_{R+1}^{r;s}-M_{R+1}^{r+1;s}\right)  \tag{8.1}
\end{equation}
Consider now the left hand side of (7.1) and its $\mu ,\nu $ summation. As
above, the case $\mu \cap \nu \equiv \alpha \cap \beta =r+1$ will give rise
to the bulk contribution $\left( M_{\widehat{r+2};\widehat{r+2}}^{r+1;r+1}%
\text{ in the previous case, see (6.3, 6.8)}\right) $ plus a remainder which
combined with the contribution for $\mu \cap \nu =r$ will construct a term
of order $1/R$ when the continuity condition for the RFT's ((5.10) and
alike) is taken into account.\ This we now exhibit first when $s<r.$

The chosen structure of the $\alpha \beta \gamma $ tree is as

\begin{equation}
\begin{array}{l}
\alpha  \\ 
\beta  \\ 
\gamma 
\end{array}
\left[ 
\begin{array}{llll}
& a_{s}...a_{r-1} & a_{r} & 0_{r+1}...a_{R} \\ 
a_{{\rm o}}...a_{s-1} & a_{s}...a_{r-1} & a_{r} & b_{r+1}^{\prime }...b_{R}
\\ 
& c_{s}...c_{r-1} & c_{r} & c_{r+1}...c_{R}
\end{array}
\right]   \tag{8.2}
\end{equation}
with $c_{s}\neq a_{s}\left( \alpha \cap \gamma =\beta \cap \gamma =s\right) $
and $b_{r+1}^{\prime }\neq 0_{r+1}\,\left( \alpha \cap \beta =r+1\right) .$

Consider first the configurations where the pair $\mu ,\nu $ is squatting
the $\alpha ,\beta $ branches of the tree

$\qquad \qquad \qquad \qquad ${\it (i)}${\it \;}\mu \cap \nu =\alpha \cap
\beta =r+1$ as in (5.5), or

$\qquad \qquad \qquad \qquad ${\it (ii)}${\it \;}\mu \cap \nu =r\neq \alpha
\cap \beta $ as in (5.6)

In that case the $\mu ,\nu $ sum yields terms in strict correspondence with
(6.3, 5) i.e., pulling them together, with the left hand side of (6.8):

\begin{equation}
\left[ _{R}W_{\widehat{r+2};\widehat{r+2};R+1}^{r+1;s;r+1}+\frac{p_{r+2}}{8}%
\left( _{A}W_{\widehat{r+2};R+1}^{r+1;s;r+1}-_{A}W_{\widehat{r+2}%
;R+1}^{r+1;s;r}\right) \right] \left( q_{r}-q_{r+1}\right)   \tag{8.3}
\end{equation}
Note that the RFT $_{R}W_{\widehat{r+2};\widehat{r+2};R+1}^{r+1;s;r+1}$ is
calculated from $W_{u;v;R+1}^{r+1;s,r+1}$ as in (4.2) with weights $%
p_{u},p_{v}.$ But $_{A}W_{\widehat{r+2};R+1}^{r+1;s;q},$ with $q=r,r+1,$ is
defined from $_{A}W_{t;R+1}^{r+1;s;q}$ with weights $p_{t}^{\left(
r+1,s,q\right) }$ instead of $p_{t}^{\left( r+1,q\right) }$ (as of (4.5) for 
$_{A}M^{r+1,q}).$ We thus get a factor $1/8$ in (8.3) (instead of $1/4$ in
(6.8)).

Consider now the configurations where $\mu ,\nu $ sit upon $\gamma $
branches, that is

(i) $\mu \cap \nu =r+1,$ i.e. (with as in (8.2) $c_{s}\neq a_{s}),$

\begin{equation}
\begin{array}{l}
\mu \\ 
\\ 
\nu
\end{array}
\left[ 
\begin{array}{llll}
&  & \quad \quad \;\;c_{r} & 0_{r+1}...m_{R} \\ 
a_{{\rm o}}...a_{s-1} & c_{s}...c_{r-1} &  &  \\ 
&  & \quad \quad \;\;c_{r}\quad \quad & n_{r+1}^{\prime }...n_{R}
\end{array}
\right]  \tag{8.4}
\end{equation}

(ii)$\mu \cap \nu =r$

\begin{equation}
\begin{array}{l}
\mu \\ 
\\ 
\nu
\end{array}
\left[ 
\begin{array}{llll}
&  & \quad -1+c_{r}\quad \quad & 0_{r+1}...m_{R} \\ 
a_{{\rm o}}...a_{s-1} & c_{s}...c_{r-1} &  &  \\ 
&  & \quad \qquad c_{r}\quad \quad & n_{r+1}^{\prime }...n_{R}
\end{array}
\right]  \tag{8.5}
\end{equation}
generating a contribution

\begin{equation}
_{\dfrac{p_{r+2}}{8}}\left( _{A}W_{\widehat{r+2};R+1}^{r+1;s,r+1}-_{A}W_{%
\widehat{r+1};R+1}^{r+1;s,r}\right) \left( q_{r}-q_{r+1}\right)  \tag{8.6}
\end{equation}
Pulling together (8.1, 3, 6) we get, for $s<r$

\begin{equation}
_{R}W_{\widehat{r+2},\widehat{r+2};R+1}^{r+1;s;r+1}=\left(
M_{R+1}^{r;s}-M_{R+1}^{r+1;s}\right) /\left( q_{r}-q_{r+1}\right) +O\left(
1/R\right)  \tag{8.7}
\end{equation}
after taking account of the continuity of RFT's.

\section{Relating 3-point and 2-point functions: $s>r+1$}

The starting tree is now, for $s>r+1,$

\begin{equation}
\begin{array}{l}
\alpha \\ 
\beta \\ 
\gamma
\end{array}
\left[ 
\begin{array}{lllll}
& a_{r} & 0_{r+1} & a_{r+2}........ & ......a_{R} \\ 
a_{{\rm o}}...a_{r-1} & a_{r} & b_{r+1}^{\prime } & b_{r+2}...b_{s-1} & 
b_{s}...b_{R} \\ 
& a_{r} & b_{r+1}^{\prime } & b_{r+2}...b_{s-1} & c_{s}...c_{R}
\end{array}
\right]  \tag{9.1}
\end{equation}
where $a\cap \beta =r+1$ (with $b_{r+1}^{\prime }\neq 0_{r+1})$ and $\beta
\cap \gamma =s,$ with $b_{s}\neq c_{s}.$

As above, the case $\mu \cap \nu \equiv \alpha \cap \beta =r+1$ will give
rise to the bulk contribution (plus a remainder). The difference is now,
with $\beta \cap \gamma =s>r+1,$ that there is no separate squatting of the $%
\gamma $ branches, as in (8.4--6).\ Instead we get only a contribution like
(8.3)

\begin{equation}
\left[ _{R}W_{\widehat{r+2};\widehat{r+2}^{\left( s\right)
};R+1}^{r+1;s:r+1}+\frac{p_{r+2}}{8}\left( _{A}W_{\widehat{r+2}%
;R+1}^{r+1;s;r+1}-_{A}W_{\widehat{r+2};R+1}^{r+1;s;r}\right) \right] \left(
q_{r}-q_{r+1}\right)  \tag{9.2}
\end{equation}
but with a difference. As noted by the $s$ superscript on the hatted
variable, the RFT of $_{R}W$ is calculated from $W_{u;v;R+1}^{r+1;s;r+1}$
with $p_{u},p_{v}^{\left( s\right) }$ respectively as in (7.3--5). The
squatting of the $\gamma $ branches by replica $\nu $ (or $\mu ),$ as in
(8.6) for $s<r,$ shows up here by the occurrence of a jump in $p_{v}^{\left(
s\right) }$ when $v\,$crosses $s.$ So that, we may alternatively separate
out a $_{R}W$ regular, that keeps the same RFT definition (with $p_{u},p_{v}$
as for $_{R}M$ in (4.2, 3)) valid for all $s,$ and exhibit the jump by
writing

\begin{equation}
s<r\;\;\qquad \qquad \quad _{R}W_{\widehat{r+2};\widehat{r+2}%
;R+1}^{r+1;s;r+1}\quad \qquad \qquad \qquad \qquad \qquad  \tag{9.3}
\end{equation}
and\footnote[5]{%
i.e. $_{AR\!}W+_{NR\!}\!W$ as hinted in footnote 4.}

\begin{equation}
s>r+1\qquad _{R}W_{\widehat{r+2};\widehat{r+2}^{\left( s\right)
};R+1}^{r+1;s;r+1}\equiv \quad _{R}\!W_{\widehat{r+2};\widehat{r+2}%
;R+1}^{r+1;s;r+1}+_{R}\!W_{\widehat{r+2};\widehat{s+1};R+1}^{r+1;s;r+1} 
\tag{9.4}
\end{equation}

Note, also, that the {\it difference} of RFT's as occuring in (9.2) is
precisely the combination of $O\left( 1/R\right) $ that a choice $\delta
P_{\alpha \beta \gamma }^{\left( r\right) }=0$ would have exhibited.

We may thus write, pulling together (8.1) and (9.2--4), for the case $s>r+1$

\begin{equation}
_{R}\!W_{\widehat{r+2};\widehat{r+2};R+1}^{r+1;s;r+1}+_{R}\!W_{\widehat{r+2};%
\widehat{s+1};R+1}^{r+1;s;r+1}=\left( M_{R+1}^{r;s}-M_{R+1}^{r+1;s}\right)
/\left( q_{r}-q_{r+1}\right) +O\left( 1/R\right)  \tag{9.5}
\end{equation}

\section{Continuity and jumps. W-T identities in the Parisi limit}

We comment first on continuity before returning to the above
identities. Consider the RFT's of some function $f$ (on its cross-overlaps,
the only RFT's considered here). From its definition, we have

\begin{equation}
p_{u}\left( f_{u}-f_{u-1}\right) =f_{\widehat{u}}-f_{\widehat{u+1}} 
\tag{10.1}
\end{equation}
Hence, at the upper bound $\left( p_{R+1}\equiv 1\right) ,$ we get

\begin{equation}
f_{R+1}-f_{R}=f_{\widehat{R+1}}  \tag{10.2}
\end{equation}
This implies a jump of $f_{u}$ at $u=R,$ if $f_{\widehat{R+1}}$ is non-zero.
This is what is seen, e.g., on the bare propagators$^{\text{[8]}}$ in the
Parisi limit, between the plateau value $x=x_{1},$ and the maximal value $%
x=1.$

At the other end of the summation domain, we get

\begin{equation}
p_{{\rm o}}f_{{\rm o}}=f_{\widehat{{\rm o}}}-f_{\widehat{1}}.  \tag{10.3}
\end{equation}
Here, even with a finite $f_{{\rm o}},$ we would get no jump for $f_{%
\widehat{{\rm o}}},$ since $p_{{\rm o}}\equiv n.$ Likewise at the lower
bound $r+1$ of the summation domain, for $f_{u}^{r}$ with $u\geqslant r+1$
as in a Replicon geometry, we have

\begin{equation}
p_{r+1}f_{r+1}^{r}=f_{\widehat{r+1}}^{r}-f_{\widehat{r+2}}^{r}  \tag{10.4}
\end{equation}

Again $f_{\widehat{r+1}}^{r}$ is found continuous and $f_{r+1}^{r}$
vanishing in the Parisi limit e.g. for the bare propagators$^{\text{[8]}}$.
Finally, in the presence of passive overlaps $r,s,$ we have

\begin{equation}
p_{u}^{\left( r;s\right) }\left( f_{u}^{r;s}-f_{u-1}^{r;s}\right) =f_{%
\widehat{u}}^{r;s}-f_{\widehat{u+1}}^{r;s}  \tag{10.5}
\end{equation}
with $p_{u}^{\left( r;s\right) }$ as of (4.7).\ Here too the RFT
(difference) is found regular, whereas $f_{u}^{r;s}-f_{u-1}^{r;s}$ has a
jump when $s$ and $u$ cross each other, compensating for the jump of $%
p_{u}^{\left( r;s\right) }.$ On the other hand eq.(5.10) shows the
continuity of the RFT under change of one of its passive overlaps.

Let us consider now equations (8.7) and (9.5) that give relationships
between the (derivative of the) mass operator $M_{R+1}^{r;s}$ and the
3-point function RFT. Letting

\begin{equation}
x=r/R+1,\;y=s/R+1,\;R\rightarrow \infty  \tag{10.6}
\end{equation}
we first get from (8.7), for $0\leqslant y<x<x_{1}$

\begin{equation}
_{R}W_{\widehat{x+{\rm o}};\widehat{x+{\rm o}};1}^{x;y;x}=\frac{\partial }{%
\partial x}M_{1}^{x;y}/\dot{q}\left( x\right) ,\qquad \qquad y<x  \tag{10.7}
\end{equation}
Here we have a {\it double} RFT, in a Replicon-geometry, the Parisi limit of
a RFT being given in this geometry by$^{\text{[14, 15]}}$

\begin{equation}
f_{\hat{k}}^{x}=\int_{k}^{x_{1}}u\;du\;\left( \partial f_{u}^{x}/\partial
u\right) +f_{1}^{x}-f_{x_{1}}^{x}\qquad \qquad k>x  \tag{10.8}
\end{equation}
In the situation (9.5), we obtain instead a left hand side with a new term $%
_{R}W_{\widehat{r+2};\widehat{s+1};R+1}^{r+1;s;r+1},$ {\it defined only for }%
$s>r+1.$ With (10.6) and for $0\leqslant x<y<x_{1}$ we get

\begin{equation}
_{R}W_{\widehat{x+{\rm o}};\widehat{x+{\rm o}};1}^{s;y;x}+_{R}W_{\widehat{x+%
{\rm o}};\widehat{y+{\rm o}};1}^{s;y;x}=\frac{\partial }{\partial x}%
M_{1}^{x;y}/\dot{q}\left( x\right) ,\qquad \qquad x<y.  \tag{10.9}
\end{equation}
The derivative of the 2-point function is thus shown to have a jump, when
the overlaps $x,y$ cross each other

\begin{equation}
\left. _{R}W_{\widehat{x+{\rm o}};\widehat{y+{\rm o}};1}^{x;y;x}\right|
_{y=x+{\rm o}}=\frac{\partial }{\partial x}M_{1}^{x;y}/\dot{q}\left(
x\right) \left| 
\begin{array}{l}
y=x+{\rm o} \\ 
\\ 
y=x-{\rm o}
\end{array}
\right.  \tag{10.10}
\end{equation}

Note that this behavior is easily checked at the zero loop level where one
keeps only the term $wtr\;q^{3}/3!$ in the free energy functional. The left
hand side (third derivative) is only non-vanishing when the ultrametric
inequality $x<y$ is satisfied, and yields $w\Theta \left( y-x\right) $ as a
result. The right hand side is the $x$ derivative of $wq$ $\left( \min
\left( x,y\right) \right) /\dot{q}\left( x\right) ,$ thus verifying
equations (10.7, 9).

\section{In conclusion:}

Let us summarize what has been accomplished. Following the approach of
(I) we have first spelled out in detail the derivation of results given in
(I). We have recalled how to build a ``small permutation'' allowing to
obtain W-T relationships to order $1/R.$ Then we have used it to derive an
equation expressing the continuity of the RFT for the 2-point function.
Secondly we have given the derivation of the identity exhibiting Goldstone
modes (and massive modes when the magnetic field is overlap dependent).
Finally we have derived and spelled out typical W-T identities relating
3-point functions to an overlap derivative of 2-point functions (in a case
where it remains rather easily tractable).

Obviously, with enough patience, the above technique may be used to
construct a complete set of W-T identities. What remains to be seen is
whether this new tool will perform the job one is usually expecting from
such identities.\ In other words, the question is now whether these W-T
identities will help to control the proliferating infrared divergences that
plague the computation of loop corrections in the condensed spin-glass
phase. But this is another story.

\section{Aknowledgments:}
Support from the Hungarian Science Fund (OTKA, \# T017493, T019422)
and from the French Ministry of Foreign Affairs (CEA/MAE \# 43) is
gratefully acknowledged.

\end{document}